\documentclass[useAMS,usenatbib,usegraphicx]{mn2e}

\def \GRB{GRB~081203A}
\def \zo{zeroth order}
\def \uvg{UV-grism}

\title[\GRB: {\it Swift}-UVOT captures the earliest UV GRB spectrum]
  {\GRB : {\it\bf Swift}-UVOT captures the earliest ultraviolet spectrum of a Gamma Ray Burst} 

\author[The Swift UVOT Team]
  {N.~P.~M. Kuin$^1$\thanks{email: npmk@mssl.ucl.ac.uk}, 
                   W. Landsman$^2$,   M.~J. Page$^1$, P. Schady$^1$, M. Still$^1$, 
     \newauthor
      A.~A. Breeveld$^1$, M. De  Pasquale$^1$, P.~W.~A. Roming$^3$, P.~J. Brown$^3$,     
     \newauthor  
       M. Carter$^1$, C. James$^1$, P.~A. Curran$^1$, A. Cucchiara$^3$, C. Gronwall$^3$,     
      \newauthor
      S.~T. Holland$^2$, E.~A. Hoversten$^3$, S. Hunsberger$^3$, T. Kennedy$^1$,S. Koch$^3$,   
     \newauthor
      H. Lamoureux$^1$, F.~E. Marshall$^2$, S.~R. Oates$^1$, A. Parsons$^2$, D.~M. Palmer$^4$, 
      \newauthor 
      and P.~J. Smith$^1$\\  
$^1$Mullard Space Science Laboratory/UCL, Holmbury St. Mary, Dorking, Surrey, RH5 6NT, UK\\
$^2$NASA/Goddard Space Flight Center, Greenbelt, MD  20771, USA\\
$^3$Department of Astronomy \& Astrophysics, Penn State University, 525 Davey Laboratory, University Park, PA 16802,
USA\\
$^4$Los Alamos National Laboratory, P.O. Box 1663, Los Alamos, NM 87545}

\begin{document}
\label{firstpage}

\date{Accepted Received: Revision 1}
\pagerange{\pageref{firstpage}--\pageref{lastpage}}
\pubyear{2008}
\maketitle

\begin{abstract}

We present the earliest ultraviolet spectrum of a gamma-ray burst
(GRB) as observed with the $\it Swift$-UVOT.  The \GRB\ spectrum was
observed for 50 seconds with the \uvg\ starting 251 seconds after the
$\it Swift$-BAT trigger. During this time the GRB was $\approx 13.4$ 
mag ($u$-filter) and was still rising to its peak optical brightness.
In the \uvg\ spectrum we find a damped Ly-$\alpha$ line, Ly-$\beta$, and
the Lyman continuum break at a redshift $z = 2.05 \pm 0.01$.  A model
fit to the Lyman absorption implies a gas column density of 
log $N_{\hbox{HI}} = 22.0 \pm 0.1$~cm$^{-2}$, 
which is typical of GRB host galaxies with
damped Ly-$\alpha$ absorbers.
This observation of \GRB\ demonstrates that for brighter GRBs 
($v \approx 14$ mag) with moderate redshift ($0.5 < z < 3.5$) 
the UVOT is able to
provide redshifts, and probe for damped Ly-$\alpha$ absorbers within
4-6 minutes from the time of the $\it Swift$-BAT trigger.
   
\end{abstract}


\section{Introduction}
\label{sec1}
Gamma-ray bursts (GRBs) are the most luminous cosmic explosions known, 
and their afterglows extend throughout the electromagnetic spectrum, 
from X-ray to radio wavelengths. Spectra of GRB afterglows in the 
ultraviolet (UV) to optical wavelength range are particularly important 
for the study of GRBs, because the spectral positions of absorption lines from the surrounding 
environment provide the redshifts, and hence the distances and 
luminosities of GRBs.
Under exceptionally favourable conditions, prompt, very near-UV 
GRB spectra have been obtained from the ground (i.e. {\it VLT}-UVES, 
which in principle is sensitive to wavelengths longward of 3000~\AA, 
observed a GRB in $\approx 10$ minutes). However, such observations are 
rare due to atmospheric limitations, time of occurrence, and the source position in the sky. 
Shorter wavelength UV GRB spectra have so far only been 
taken with the {\it HST}/STIS \citep{Smette}, at least 3 days after the GRB trigger. 
In this letter we present the first prompt UV spectrum 
of a GRB. The spectrum was obtained with the \uvg\ of the {\it Swift} 
Ultra-Violet/Optical Telescope \citep[UVOT;][]{Roming}.

Since the launch of the {\it Swift} satellite \citep{swift} in 2004, 
the UVOT has been making prompt photometric 
observations of GRBs in the optical and UV, simultaneously with the X-ray and 
gamma-ray observations taken by the X-Ray Telescope \citep[XRT;][]{Burrows} and 
the Burst Alert Telescope \citep[BAT;][]{Barthelmy}, respectively. The UVOT 
photometric system \citep{Poole} provides photometry in three UV and three
optical bands, and in one $white$-filter which is sensitive 
over the range of 1700--8000~\AA. 

Two grisms are also included in the UVOT filter wheel to enable spectra of GRBs to be obtained.
Until recently they have not been used in the GRB observing sequence 
because early in the mission the data indicated that most 
GRB afterglows were too faint for grism observations, even at early times \citep{Roming2006}.
However, after four years of {\it Swift} observations, our GRB sample
is large enough that we can meaningfully reassess the frequency of
GRB afterglows which are bright enough for grism spectroscopy, and ascertain 
the optimum timing for their inclusion in the automated observing sequence.
The catalog of GRB afterglows observed with the UVOT in the first two-and-a-half years 
\citep{Roming2008} shows that UVOT observes $\approx 90$ each year, of 
which $\approx 25$ are detected by UVOT in at least one exposure. In its first four years of 
operation, {\it Swift} observed 5 GRBs (GRB~050525A, GRB~050922C, 
GRB~061007, GRB~080319B and GRB~080810) 
brighter than 14th magnitude in $v$. It is GRBs like these that should provide 
good spectra through the UVOT grisms. 

The \uvg\ was introduced into the automated UVOT GRB response on the 7th Oct, 2008. 
In this latest sequence, the grism spectrum is 50 seconds long, proceeded by 
an initial 150 second $white$-filter finding chart 
and followed by a second finding chart in the $u$ filter. With a  median 
{\it Swift} slew time of 86 seconds \citep{Roming2008}, the grism observation 
typically begins 225-275 seconds after the burst trigger, and covers the 
wavelength range from $\approx$~1700-5800~\AA. The onboard software 
autonomously replaces the grism exposure with a $u$-filter exposure 
for bursts which are faint in the BAT, and therefore unlikely to be viable grism targets.  

In the following sections we present the first grism image of a GRB taken with the UVOT 
\uvg, the resulting spectrum of \GRB, and its corresponding analysis.

\begin{figure}
\includegraphics[width=88.0mm,angle=0]{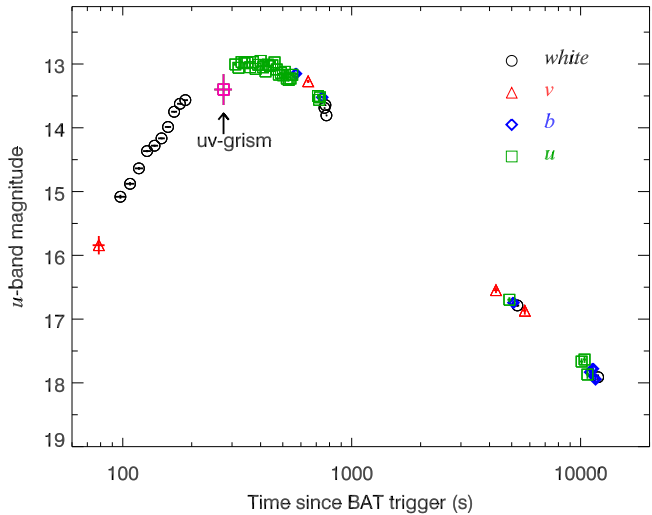}
\caption{The lightcurve of \GRB\  shows that the spectrum was taken during the rise. 
The photometry point derived
from the spectrum, indicated with an arrow, 
has a larger uncertainty than those of the lenticular filters.}
\label{fig2}
\end{figure}

\section{Observations and data reduction}    
\label{sec2}
The BAT instrument triggered on \GRB\ at $T$ = 2008-12-03 at 13:57:11 UT \citep{batcirc}.  
The mask-weighted light curve showed two overlapping peaks: the first starts
at $T$-69 seconds and peaks at $T$+10 seconds, the second peaks at $T$+32 seconds and 
ends at $T$+405 seconds. 
UVOT started settled observations of \GRB\ 93 seconds after the trigger beginning with a 150
second $white$-filter finding chart, followed by a 50 second \uvg\
exposure starting 251 seconds after the trigger, a second finding chart
in $u$, and then exposures in the other UVOT photometric filters \citep{uvotcirc}.
The \uvg\  spectrum was taken during the decay of the second BAT gamma-ray peak. 
The best source position was determined by the UVOT, which
located a fading source at 
${\rm RA_{J2000}}$ = 15h 32m 07.58s (=233.03158$\deg$), ${\rm Dec_{J2000}}$ 
= +63d 31m 14.9s (=63.52081$\deg$), with an uncertainty of 0.5\arcsec 
\citep[90\% confidence limit; ][]{uvotcirc}, 
consistent with the refined XRT position \citep{xrtrefpos}. 
Using the grism spectrum, a provisional redshift of 2.1 was reported by \citet{redshift}. 
The X-ray spectrum, extracted from the XRT Window Timing data, overlaps with the 
grism exposure and has an exposure time of 558 s, starting 87 seconds after the 
trigger \citep{xrtcirc}. 

The source was detected in the $u,b,v$, $white$, $uvw1$ filters, and only faintly in the $uvw2$ filter. It 
was not detected in the $uvm2$ filter, which is consistent with the filter response curves and
the shape of the \uvg\ spectrum.  The brightness of the GRB can be
measured from the \zo\ of the grism image, which provides a
magnitude of $b=13.76\pm 0.23$ at the time of the grism exposure. 
Folding the extracted spectrum (see Section
\ref{sec3}) through the $u$ filter response leads to a magnitude of $13.4$ mag
during the grism exposure with a systematic error estimated to be 25\%. 
The light curve of \GRB\ is shown in Figure \ref{fig2}. 

\begin{figure}
\includegraphics[width=86mm]{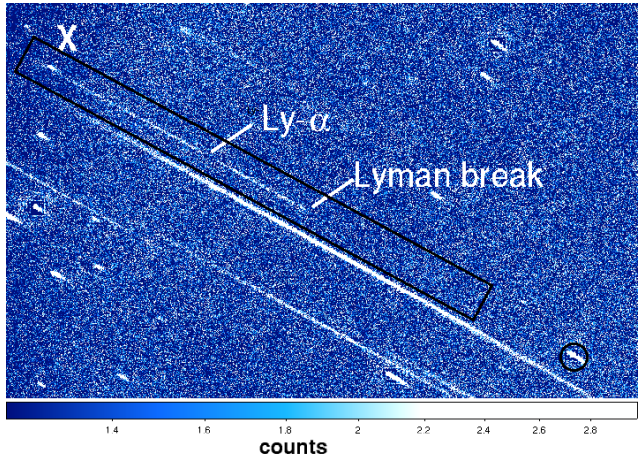} 
\caption{The \uvg\ image of \GRB. The GRB first order spectrum is inside the black rectangle.
The \zo\ is indicated by a black circle. The `X' in the top left corner 
indicates the contaminating \zo\ of a star.  }
\label{fig1}
\end{figure}

A cutout of the grism image with the GRB spectrum is shown in Figure \ref{fig1}.  
The \zo\ of a 15th magnitude star, 
falls on the GRB spectrum around 5000~\AA. 
The image was processed using the {\sc HEADAS-6.5.1} software to remove the {\sc modulo-8} 
pattern \citep{Poole}. 

The raw count rate spectrum was extracted using the {\sc uvotimgrism} program. 
The data were binned up by a factor of three, which reduces
any correlations caused by the image rotation during the spectral extraction but keeps the 
bin size smaller than the wavelength resolution. 
The wavelength scale was adjusted to the latest \uvg\ wavelength calibration
(to be released shortly)
for the nominal filter wheel position, and is accurate to 15~\AA.
The \uvg\ spectrum has a resolution of $R \approx 150$ at 2600\AA, with 
a dispersion of $\sim 3.2$~\AA/pixel.
It can be seen that blueward of 2780\AA\ no signal above the background is seen.
Although wavelengths in the first order spectrum longer than 2850~\AA\ can be 
contaminated by a second order spectra, this is not a concern here since there 
is no source flux shortward of 2780~\AA, thus, the second order spectrum 
will only affect the first order for wavelengths longer than 5549~\AA.  

\section{Spectral modelling}
\label{sec3}

\begin{figure}
\includegraphics[width=59mm,angle=-90]{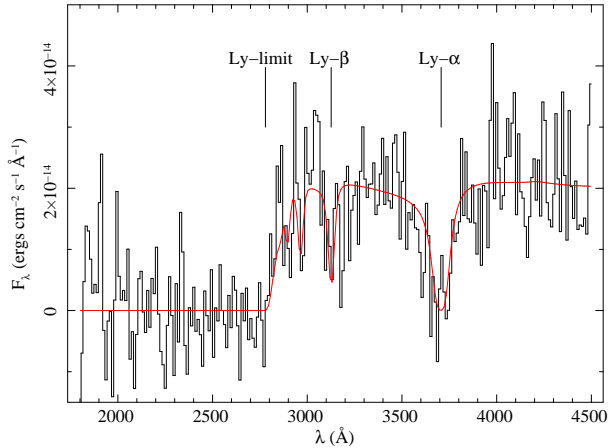} 
\caption{The \uvg\ first order spectrum of \GRB. 
The model shown here as a red line includes the Lyman forest as discussed 
in Section \ref{sec3}.}
\label{fig3}
\end{figure}

The background subtracted spectrum
of \GRB\ is provided in Figure \ref{fig3}, overlayed by our best-fit
spectral model. Calibrations for the wavelength
scale, line spread function, and telescope throughput have been
extracted from a development version of the Swift calibration data
base. Wavelengths are currently good to $0.5\%$ systematic accuracy
($1\sigma$) and the flux calibration is good to an estimate of $25\%$.

Neglecting the systematic uncertainty in the wavelength scale, 
Figure \ref{fig3} reveals a broad absorption line at
3706$\pm$11\AA\ that we identify with the Ly$\alpha$ transition of
hydrogen, and a corresponding Lyman continuum edge at 2779$\pm$8\AA,
shortward of which no significant source detection is
made. Ly$\beta$ is weakly detected at 3126$\pm$8\AA. 
Wavelength errors and the best-fit parameters are at 
the $1\sigma$ level. 

Within the spectral model, the afterglow continuum is characterised by
a single power law attenuated by neutral gas and dust from both the
Milky Way and host galaxy of the burst.  Both galactic and host galaxy
continuum absorption at wavelengths shorter than the Lyman edge are
modelled using the photo-electric cross-sections of \citet{morrison}
and the relative element abundances of \citet{anders}.
The Milky Way extinction law has been taken from the
analytical description of \citet{pei},
and the host extinction is
assumed identical to the Small Magellanic Cloud law, also from \citet{pei}.
Host galaxy hydrogen Lyman series absorption from $n$ = 1--100
is modelled following the algorithm described in \citet{totani}
and based upon \citet{peebles}.
Oscillator strengths for transitions n= 1--31 are taken from \citet{morton}
and extrapolated to n $\leq$ 100 thereafter. 
The wavelength-dependent optical depth
calculated within the model is also broadened by a macroscopic
velocity field within the absorbing column. We assume a Gaussian
velocity field, characterised by its full-width half-maximum (FWHM)
velocity.

The combined model contains four free parameters: cosmological
redshift ($z$), neutral hydrogen column density ($N_{\mbox{HI}}$), FWHM of the
absorbers velocity field ($v$), and the normalization of the afterglow
continuum. In order to constrain the continuum emission over the
relatively narrow spectral range of the \uvg, five additional
model parameters are pre-determined.  The Galactic column density
towards the burst, $\log{N_{\mbox{H,g}}}$ = 20.23, is taken from the
map of \citet{kalberla}. Galactic dust is quantified by the extinction 
coefficient $A_{\mbox{V,g}}$ = 0.06, taken from \citet{schlegel}.
The continuum spectral index ($\beta$ = 0.90$\pm$0.01), 
the neutral column density within the host galaxy
($\log{N_{\mbox{H,h}}} = 21.7\pm 0.1$), 
and the host galaxy dust extinction coefficient ($A_{\mbox{V,h}}$ = 0.08)  
were all measured by a fit to the broad band spectral energy distribution (SED)
of the afterglow, interpolated to the $T$+700s epoch, and incorporating six lenticular
filter magnitudes from the UVOT and the X-ray spectrum from the
XRT. This method is described by \citet{Schady}.
The detailed fit to the SED will be provided in a forthcoming paper concerning the
spectral and temporal behaviour of \GRB.
 
The model fit to the grism spectrum was performed using {\sc XSPEC} 12.5.0 \citep{arnaud}.
The best-fit provides $\chi^2$ = 301.9 for 231 degrees of
freedom (dof), $\log{N_{\mbox{HI}}}$ = 22.09$^{+0.09}_{-0.12}$, $z$ =
2.046$ \pm 0.002$, and $v <$ 598~km/s (90\% confidence limit). 
The significance of the line detections was determined
by freezing the redshift. The fit to the spectrum 
in the 3400 -- 4000~\AA\ wavelength range gives a $4.0\sigma$ 
detection of Ly$\alpha$, and $Ly\beta$ is detected at the $1.5\sigma$ 
level in the 3000 -- 3300~\AA\ wavelength range.
The best-fit column density satisfies the definition for a damped Ly$\alpha$
absorption system (DLA). Taking into account systematic errors in the
wavelength dispersion, $z$ =2.05$\pm$0.01.

At a redshift of 2.05, we expect some contamination in the spectrum
from absorption systems located between the Earth and the host \citep{Madau}.
Within the limits of the signal-to-noise, there is no compelling
evidence for discrete Ly$\alpha$ absorption from intervening galaxies. 
We do however investigate
whether an undetected, or unresolved, Lyman forest could have an
impact on the measured properties. We add the statistically-based
algorithm characterising the cosmological hydrogen density from \citet{Madau}
to the model, which effectively absorbs photons blueward of the 
host galaxy Ly-$\alpha$ line (Figure \ref{fig3}). The
best fit for this situation provides $\chi^2$ = 271.4 for 231 dof, 
$\log{N_{\mbox{HI}}}$ = 22.02$^{+0.11}_{-0.12}$, $z$ =
2.05$\pm$0.01, and $v <$ 474~km/s. The Madau model gives a slightly 
better fit to the data.

\section {Discussion and Conclusions}
\label{sec4}

\begin{figure}
\includegraphics[width=84mm]{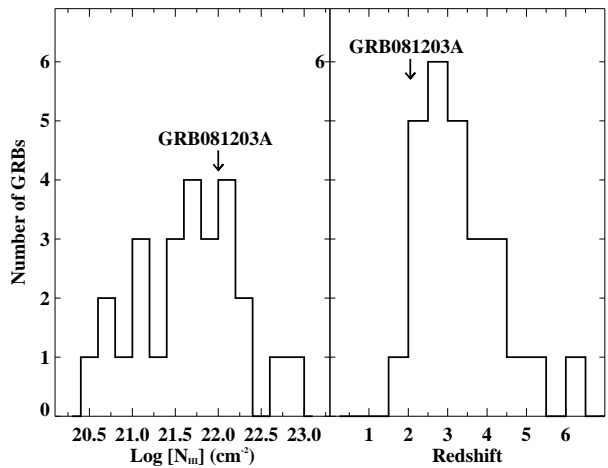} 
\caption{The figure shows the number of GRBs with an associated damped Lyman-$\alpha$ system (DLA). 
The left panel shows that the $N_{HI}$-column density of \GRB\ as determined in this 
paper is quite typical. The right-hand panel shows that the redshift of \GRB\   
is one of the lowest measured for a GRB with a DLA. The UVOT is capable of 
measuring redshifts below $z = 1.9$ that cannot be measured from the ground. 
The data for this figure were taken from the literature \citep{prochaska2007,jacobson,savaglio}.}
\label{fig4}
\end{figure}

With the rapid and precise locations of GRBs available with {\it Swift}, the 
number of GRBs available with good signal-to-noise spectrosopic data has doubled. 
Of these, $16\%$ have an associated DLA system \citep{prochaska2007, jacobson}.
The column density of $N_{HI} = 1.04\times 10^{22}$~cm$^{-2}$ in \GRB\ is  
typical of GRB-DLAs \citep{prochaska}, but its 
redshift is amongst the smallest of the sample (see Figure \ref{fig4}). 
It also has the lowest redshifts observed in a GRB for which both the Lyman break and 
Ly-$\alpha$ have been obtained. 

Since it is possible for UVOT spectra to be taken during the rising phase of the 
afterglow, and while the gamma-ray emission is still in progress, as was the case
for \GRB, useful constraints on the evolution of the ionization state of the gas in the GRB environment 
may be obtained by comparison to 
ground-based spectra taken later in the decay phase of the afterglow.
Getting an early spectrum with UVOT opens up the opportunity to increase the 
sample of {\it Swift} GRBs with redshifts. 
As well as its obvious advantage for UV coverage, the UVOT is capable of taking spectra 
of GRBs with positions that are unobtainable for ground based observations. 
Indeed, \GRB\ was a twilight object for ground based observers, and we 
are not aware of any ground-based spectroscopic follow-up of this object. 

We have shown that it is possible to obtain useful spectra using the UVOT \uvg\  in the 
automated {\it Swift} response to new GRBs. The combined measurement of the Ly-continuum edge, 
Ly-$\alpha$, and possibly Ly-$\beta$ provides 
a good measurement of the redshift in the UVOT \uvg\ spectra for redshifts of GRBs 
between $0.5 < z < 3.5$. Using such spectra, we expect to find new DLAs in the 
range $0.5 < z < 2.0$. 
$Swift$ observations can in principle probe the different components of the interstellar medium 
by providing a measurement of $HI$ from the grism spectrum, 
dust extinction from the photometry, and metallicity from the X-rays.

\section*{acknowledgements}
This work was supported by the UK Science and Technology Facilities
Council through a grant for {\it Swift} Post Launch Support at
UCL-MSSL. This work is sponsored at PSU by NASA contract NAS5-00136.
We acknowledge useful comments by the anonymous referee which led to improvement of the grism analysis.
We would like to dedicate this letter to the late Richard
Bingham, whose visionary optical design for the UVOT grism made these
observations possible. 

\bibliographystyle{mn}

\bsp

\label{lastpage}

\end{document}